\documentclass{emulateapj}

\newcommand{\be}{\begin{equation}}
\newcommand{\ba}{\begin{eqnarray}}
\newcommand{\ee}{\end{equation}}
\newcommand{\ea}{\end{eqnarray}}

\begin{document}
\title{Missing Pages in Our Photo Album of the Infant Universe}

\author{Abraham Loeb\altaffilmark{1}}


\altaffiltext{1}{Astronomy Department, Harvard University, 60 
Garden Street, Cambridge, MA 02138, USA; E-mail: {\it aloeb@cfa.harvard.edu}}

\begin{abstract} 

Existing data sets include an image of the Universe when it was 0.4 million
years old (in the form of the cosmic microwave background), as well as
images of individual galaxies when the Universe was older than a billion
years.  But there is a serious challenge: in between these two epochs was a
period when the Universe was dark, stars had not yet formed, and the cosmic
microwave background no longer traced the distribution of matter.  And this
is precisely the most interesting period, when the primordial soup evolved
into the rich zoo of objects we now see.  In this popular-level overview I
describe how astronomers plan to observe this nearly-invisible yet crucial
period.

\end{abstract}

\section{\bf First Light}

When I look up into the sky at night, I often wonder whether we humans are
too preoccupied with ourselves. There is much more to the universe than
meets the eye on earth. As an astrophysicist I have the privilege of being
paid to think about it, and it puts things in perspective for me. There are
things that I would otherwise be bothered by - my own death, for
example. Everyone will die sometime, but when I see the universe as a
whole, it gives me a sense of longevity. I do not care so much about myself
as I would otherwise, because of the big picture.

Cosmologists are addressing some of the fundamental questions that people
attempted to resolve over the centuries through philosophical thinking, but
we are doing so based on systematic observation and a quantitative
methodology.  Perhaps the greatest triumph of the past century has been a
model of the universe that is supported by a large body of data. The value
of such a model to our society is sometimes underappreciated. When I open
the daily newspaper as part of my morning routine, I often see lengthy
descriptions of conflicts between people about borders, possessions or
liberties. Today's news is often forgotten a few days later. But when one
opens ancient texts that have appealed to a broad audience over a longer
period of time, such as the Bible, what does one often find in the opening
chapter? A discussion of how the constituents of the universe -- light,
stars, and life -- were created. Although humans are often caught up with
mundane problems, they are curious about the big picture. As citizens of
the universe we cannot help but wonder how the first sources of light
formed, how life came into existence and whether we are alone as
intelligent beings in this vast space. Astronomers in the 21st century are
uniquely positioned to answer these big questions with scientific
instruments and a quantitative methodology.

It is sometimes argued that science takes away the sense of mystery about
our origins. However, current scientific circumstances appear to only
enhance the mystery.  Consider what astronomers have learned so far about
the early Universe.  We have an image of the Universe at the moment that
hydrogen atoms first formed in it -- namely the cosmic microwave background
radiation -- and we have pictures of individual galaxies more than a
billion years later.  The intervening epoch, though, was a period that
started when the Universe was dark, stars had not yet formed, and the
cosmic microwave background no longer traced the distribution of
matter. And this is precisely the most interesting period, when the
primordial soup evolved into the rich zoo of objects we now see.

The situation is similar to having a photo album of a person that contains
the first ultrasound image of him or her as an unborn baby and some
additional photos as a teenager and an adult.  If you tried to guess from
these pictures what happened in the interim, you could be seriously wrong.
A child is not simply a scaled-up fetus or scaled-down adult.  The same is
with galaxies.  The primordial hydrogen gas was composed of atoms, so you
might suppose that the Universe traced out a straightforward and rather
boring path towards the assembly of atoms into galaxies.  It did not.
Observations of the spectra of early galaxies and quasars, which reveal the
conditions in their environments, indicate that the cosmic gas actually
underwent a wrenching transition from atoms back to their constituent
protons and electrons -- a process known as reionization.  In fact,
although the world around us is composed of atoms, the bulk of the
Universe's ordinary matter today is still in the form of free electrons and
protons, located deep in intergalactic space.

\begin{figure} 
\plotone{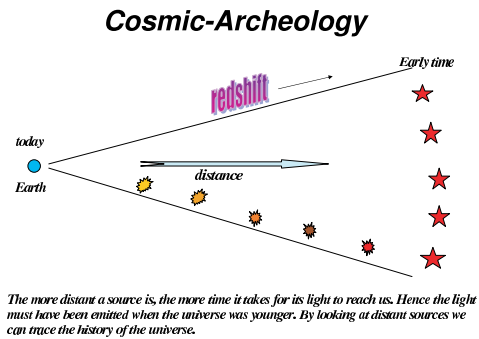} 
\caption{Cosmology is like archeology. The deeper one looks, the older is
the layer that one is revealing, owing to the finite propagation speed of
light.}
\label{fig:1}       
\end{figure}

How the Universe underwent this transition is one of the most exciting
questions in cosmology today.  Most researchers associate the transition
with the first generation of stars, whose ultraviolet radiation streamed
into intergalactic space and broke atoms apart.  Others conjecture that
material plummeting into black holes gave off radiation on its death
plunge.  But as often is the case in science, new observational data is
required to test which of these scenarios describes reality better.


The timing of reionization depends on astrophysical parameters such as the
efficiency of making stars or black holes in galaxies, but most importantly
it depends on the nature and the initial inhomogeneities of the cosmic
matter. Galaxies form by a process known as gravitational instability which
is seeded by the initial inhomogeneities. A region that starts slightly
denser than the average density of the Universe will tend to pull itself
together by its own gravity. Although initially the region expands like the
rest of the Universe, its extra gravity slows its expansion down, turns it
around, and make the region collapse upon itself to make a bound object
like a galaxy.  Most of the cosmic matter is known to be dark, i.e. have a
very weak interaction (aside from gravity) with ordinary matter and
radiation.  To get the process of galaxy formation started, one needs
inhomogeneities in the cosmic matter distribution on the small scale of
galaxies.  Such inhomogeneities exist and can seed the formation of dwarf
galaxies at early times only if the dark matter is made of massive
particles that are initially cold.  On the other hand, if the dark matter
is warm, the velocity dispersion of the dark matter particles would erase
inhomogeneities on small scales and prevent the formation of dwarf galaxies
at early times.  Existing data on the spectrum of initial inhomogeneities
favors the notion that the dark matter is cold and not warm but the
evidence is still preliminary. New data on the mass and formation time of
the first dwarf galaxies will be able to robustly determine whether this is
indeed the case.  Laboratory accelerators (such as the {\it Large Hadron
Collider}) will in parallel be able to search in the future for the
hypothetical particle that makes the cold dark matter as long as it has an
interaction strength comparable to the ``weak interaction'' of neutrinos,
as expected in some particle-physics models.

According to the popular cosmological model of cold dark matter, dwarf
galaxies started to form when the Universe was only a hundred million years
old.  Numerical simulations indicate that the first stars to have formed
out of the prestine primordial gas left over from the big bang, were much
more massive than the sun.  Lacking heavy elements which would have cooled
the gas to lower temperatures, the warm primordial gas could have only
fragmented into relatively massive clumps which condensed to make the first
stars. These stars were efficient factories of ionizing radiation (see the
Scientific American article by Bromm \& Larson).  Once they exhaust their
nuclear fuel, some of these stars are expected to explode as supernovae and
disperse the heavy elements that were cooked by nuclear reactions in their
interiors into the surrounding gas.  The heavy elements cool the diffuse
gas to lower temperatures and allow it to fragment into lower-mass clumps
that make the second generation of stars. The ultraviolet (UV) radiation
emitted by all generations of stars eventually leaked into the
intergalactic space and ionized gas far outside the boundaries of
individual galaxies.

The earliest dwarf galaxies merged and made bigger galaxies as time went
on. A present-day galaxy like our own Milky-Way was constructed over cosmic
history by the assembly of a million building blocks like the first dwarf
galaxies. The UV radiation from each galaxy created an ionized bubble in
the cosmic gas around it. These bubbles grew in size as the galaxies grew
in mass and eventually surrounded groups of galaxies. Finally as more
galaxies formed, the bubbles overlapped and the initially neutral gas in
between the galaxies was completely re-ionized.

Although the above progression of events sounds plausible, its existence
has been confined to the minds of theorists so far.  Empirical cosmologists
would like to actually see direct evidence for the reionization epoch
before putting its description as a missing chapter in their
textbooks. {\it How can one observe the reionization history of the
Universe?} ... One way is by imaging hydrogen through its radio (21cm)
emission and searching for the ionized bubbles in it with radio
telescopes. Another way is by searching for the radiation emitted by the
first galaxies with big new telescopes from the ground as well as from
space. Below we will describe each of these techniques along with the
theoretical work that motivates it.
The study of the reionization epoch promises to be one of the most active
frontiers in cosmology over the next decade.

\begin{figure}
\plotone{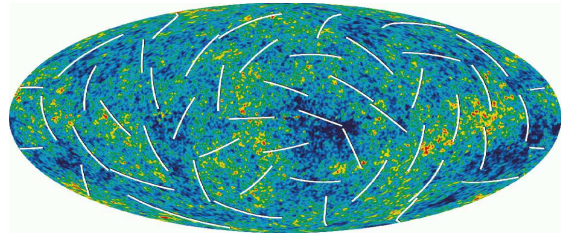} 
\caption{Image of the Universe at the time when it became transparent,
taken by the {\it WMAP} satellite (see http://map.gsfc.nasa.gov/ for
details).  The slight density inhomogeneties at the level of one part in
$\sim 10^5$ in the otherwise uniform Universe, imprinted hot and cold spots
in the brightness map of the cosmic microwave background.  The existence of
these anisotropies was predicted in a number of theoretical papers three
decades before the technology for taking this image was available.  The
white bars show the measured polarization direction of the background
light.  }
\label{WM}
\end{figure}

\section{\bf Cosmic Time Machine}

When we look at our image reflected off a mirror at a distance of 1 meter,
we see the way we looked 6 nano-seconds ago, the light travel time to the
mirror and back. If the mirror is spaced $10^{19}~{\rm cm}=3$pc away, we
will see the way we looked twenty one years ago. Light propagates at a
finite speed, and so by observing distant regions, we are able to see how
the Universe looked like in the past, a light travel time ago. The
statistical homogeneity of the Universe on large scales guarantees that
what we see far away is a fair statistical representation of the conditions
that were present in our region of the Universe a long time ago.

This fortunate situation makes cosmology an empirical science. We do not
need to guess how the Universe evolved. Using telescopes we can simply see
the way distant regions appeared within it at earlier cosmic times. Since a
greater distance means a fainter flux from a source of a fixed luminosity,
the observation of the earliest sources of light requires the development
of sensitive instruments and poses challenges to observers.

We can in principle image the Universe only if it is transparent. Earlier
than 0.4 million years after the big bang, the cosmic gas was sufficiently
hot to be fully ionized (i.e. atoms were broken into free nuclei and
electrons) and the Universe was opaque due to scattering by the dense fog
of free electrons that filled it. Thus, telescopes cannot be used to image
the infant Universe at earlier times (or redshifts $\ga 10^3$). The
earliest possible image of the Universe was recorded in the cosmic
microwave background, the thermal radiation left over from the transition
to transparency (see Fig. \ref{WM}).

\begin{figure} 
\plotone{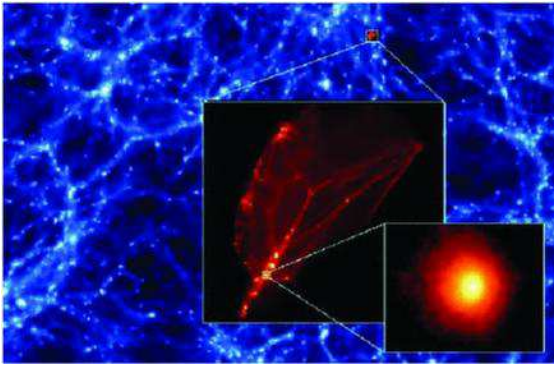} 
\caption{A slice through a numerical simulation of the first dark matter
condensations to form in the Universe. Colors represent the dark matter
density at a redshift $z=26$.  For a given redshift value $z$, the
coefficient $(1+z)$ is the factor by which any scale in the Universe
(including the wavelength of a photon) is being stretched up to the
present-time.  The simulated volume is $1.8\times 10^{20}$ present-day cm
on a side, simulated with 64 million particles each weighing $\sim 10^{-4}$
Earth masses (!). (from Diemand, Moore, \& Stadel 2005).}
\end{figure}

\section{\bf Completing Our Photo Album of the Universe}

The ultimate goal of observational cosmology is to image the entire history
of the Universe since the time it became transparent. Currently, we have a
snapshot of the Universe at an age of 0.4 million years from the microwave
background, and detailed images of its evolution starting from an age of a
billion years to the present time. The evolution between a million and a
billion years has not been imaged as of yet.

Within the next decade, NASA plans to launch the successor to the Hubble
Space Telescope named the James Webb Space Telescope ({\it JWST}) that will
be able to get infrared images of the very first sources of light (stars
and black holes) in the Universe, which are predicted theoretically to have
formed in the first hundreds of millions of years. In parallel, there are
several initiatives to construct large infrared telescopes on the ground
with the same goal in mind
\footnote{http://www.eso.org/projects/e-elt/}$^{,}$
\footnote{http://tmt.ucolick.org/}$^{,}$\footnote{http://www.gmto.org/}.
Independently, the neutral hydrogen left over from the big bang, can be
mapped in three-dimensions through its 21cm transition even before the
first galaxies formed. Several groups are currently constructing
low-frequency radio arrays in an attempt to map the initial inhomogeneities
as well the process by which the hydrogen was re-ionized by the first
galaxies.

\begin{figure} 
\plotone{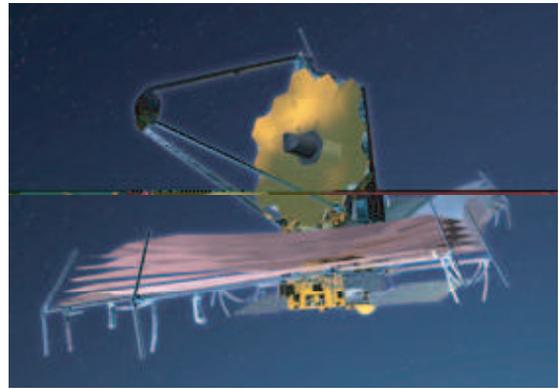} 
\caption{A sketch of the current design for the {\it James Webb Space
Telescope}, successor to the {\it Hubble Space Telescope} to be launched in
2013 (http://www.jwst.nasa.gov/). The current design includes a primary
mirror made of beryllium which is 6.5 meter in diameter as well as
instrument sensitivity that spans the full range of infrared wavelengths
0.6--28$\mu$m that will allow detection of the first galaxies.
The telescope will orbit 1.5 million km from Earth at the Lagrange L2
point.}
\end{figure}

\begin{figure} 
\includegraphics[width=0.7\columnwidth,angle=-90]{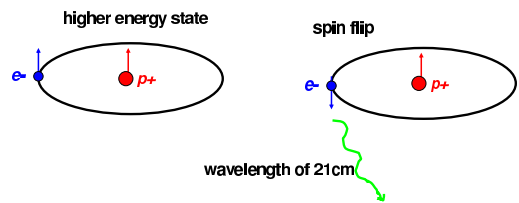} 
\caption{The 21cm transition of hydrogen. The higher energy level the spin
of the electron (e-) is aligned with that of the proton (p+).  A spin flip
results in the emission of a photon with a wavelength of 21cm (or a
frequency of 1420MHz).}
\label{21cm}
\end{figure}

\begin{figure}
\centering
\includegraphics[height=10cm,angle=-90]{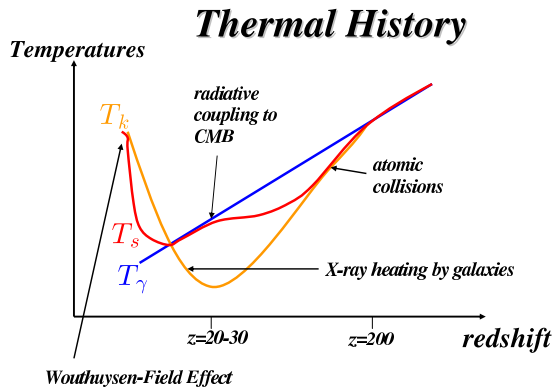}
\caption{Schematic sketch of the evolution of the kinetic and spin
temperature of cosmic hydrogen. Following cosmological recombination at a
redshift $z\sim 10^3$, the gas temperature (orange curve) tracks the Cosmic
Microwave Background (CMB) temperature (blue line; $\propto (1+z)$) down to
a redshift $z\sim 200$ and then declines below it ($\propto (1+z)^2$) until
the first X-ray sources (accreting black holes or exploding supernovae)
heat it up well above the CMB temperature. The spin temperature of the 21cm
transition (red curve) interpolates between the gas and CMB
temperatures. Initially it tracks the gas temperature through atomic
collisions; then it tracks the CMB through radiative coupling; and
eventually it tracks the gas temperature once again after the production of
a cosmic background of UV photons that redshift into the Ly$\alpha$
resonance (through the so-called Wouthuysen-Field effect [Wouthuysen
1952;Field 1959]).  }
\label{spin}
\end{figure}

\begin{figure} 
\plotone{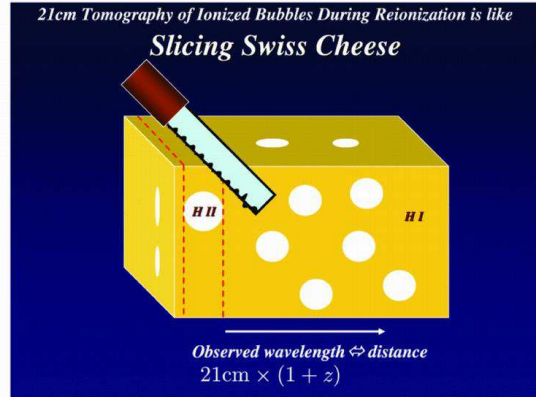} 
\caption{21cm imaging of ionized bubbles during the epoch of reionization
is analogous to slicing swiss cheese. The technique of slicing at intervals
separated by the typical dimension of a bubble is optimal for revealing
different pattens in each slice.}  \label{swiss}
\end{figure}

The wavelength of 21cm resonates with a transition of hydrogen between two
states of the electron spin (splitting hydrogen's ground state) from an
upper level in which the electron spin is aligned with that of the proton
to the lower level where it is anti-aligned with it (see Fig. \ref{21cm}).
The relative population of the two levels defines the so-called {\it spin
temperature} which may deviate from the actual kinetic temperature of the
gas in the presence of a radiation field. The coupling between the gas and
the cosmic microwave background (owing to the small residual fraction of
free electrons left over from the hydrogen formation epoch) kept the gas
temperature equal to the radiation temperature up to 10 million years after
the big bang. Subsequently, the cosmic expansion cooled the gas faster than
the radiation and collisions among the atoms maintained their spin
temperature at equilibrium with their own kinetic temperature.  At this
phase, cosmic hydrogen could be detected in {\it absorption} against the
microwave background sky since it is colder. Regions that are somewhat
denser than the mean will produce more absorption and vice versa. The
resulting fluctuations in the 21cm brightness would simply reflect the
primordial inhomogeneities in the gas.  A hundred million years after the
big bang, cosmic expansion diluted the density of the gas to the point
where the collisional coupling of the spin temperature to the the gas
became weaker than its coupling to the microwave background.  At this
stage, the spin temperature returned to equilibrium with the radiation
temperature and it is impossible to see the gas against the microwave
background brightness. Once the first galaxies lit up, they heated the gas
(mainly by emitting X-rays which penerated the thick column of
intergalactic hydrogen) as well as its spin temperature (through UV photons
which couple the spin temperature to the gas kinetic temperature). The
increase of the spin temperature beyond the microwave background
temperature requires much less energy per atom than ionization, and so this
heating occured well before the Universe was reionized. Once the spin
temperature had risen above the microwave background temperature the gas
can be seen against the microwave sky in {\it emission}. At this stage, the
hydrogen distribution is punctuated with bubbles of ionized gas which are
created around the galaxies. Mapping of the hydrogen distribution up to the
completion of reionization is expected to reveal structures similar to
those found by slicing swiss cheese (see Fig. \ref{swiss}). 

The 21cm wavelength of a photon emitted at some early cosmic time is
stretched by the cosmic expansion since that time. The wavelength observed
today is therefore larger than 21cm by a factor greater than unity which we
may express as $(1+z)$, where the excess stretch $z$ is the so-called {\it
cosmological redshift} (named after its tendency to shift blue photons with
a short wavelength towards the red where the wavelength is longer).  A
source at a greater distance requires an earlier emission time due to the
light propagation delay, providing more wavelength stretching by the
cosmic expansion, and hence a higher redshift.  We know that reionization
must have ended by the time that the Universe was a billion years old, and
so there is no point in attempting to image diffuse hydrogen at later
times. These late times correspond to redshifts below 6 or a wavelength
below $(1+6)\times 21=147$cm, corresponding roughly to the height of a
teenager.

\begin{figure}
\plotone{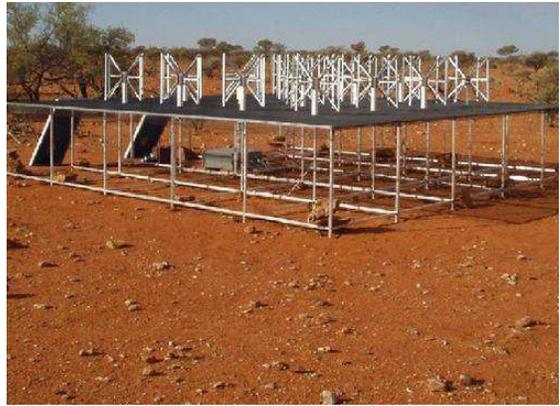}
\caption{Prototype of the tile design for the {\it Mileura Wide-Field
Array} (MWA) in western Australia, aimed at detecting redshifted 21cm from
the epoch of reionization. Each 4m$\times$4m tile contains 16 dipole
antennas operating in the frequency range of 80--300MHz. Altogether the
initial phase of MWA (the so-called ``Low-Frequency Demostrator'') will
include 500 antenna tiles with a total collecting area of 8000 m$^2$ at
150MHz, scattered across a 1.5 km region and providing an angular
resolution of a few arcminutes across the sky.}
\label{mwa}
\end{figure}

Detection of the redshifted 21cm emission by hydrogen will be made possible
with arrays of low-frequency antennas, similar to those used for television
and radio communication. By shifting in observed wavelength, these antenna
will be slicing the Universe at different redshifts, i.e. different
distances.  The combination of all the slices would provide a
three-dimensional map (i.e. tomography) of the neutral hydrogen
distribution.  Figure \ref{mwa} shows the antenna module of one of the
experiments being constructed right now, the Mileura Wide-Field Array (MWA;
http://web.haystack.mit.edu/arrays/MWA/).  Other experiments whose goal is
to detect 21cm fluctuations from the epoch of reionization at $z\sim 6-12$
include {\it LOFAR} ({\it http://www.lofar.org}), the 21CMA (formerly known
as the Primeval Structure Telescope [{\it PAST}; {\it
http://arxiv.org/abs/astro-ph/0502029}]), and in the more distant future
the Square Kilometer Array ({\it SKA}; {\it http://www.skatelescope.org}).

The 21cm mapping of cosmic hydrogen may potentially carry the largest
number of bits of information compared to any other survey method in
cosmology.  In particular, it has the potential to provide a much richer
data set than the well-established method of mapping the brightness
fluctuations of the cosmic microwave background across the sky. While the
latter provides a two-dimensional image of the surface where the microwave
photons originated (corresponding to the time when the Universe became
transparent), the redshifted 21cm photons map the hydrogen distribution in
{\it three dimensions}. These dimensions include the two sky coordinates
and the observed wavelength, which is equivalent to distance because larger
distances are associated with longer light travel time and more stretching
of the initial 21cm wavelength of the photons by the cosmic
expansion. Second, the fluctuations of the microwave background are known
to be damped on small scales, because the microwave photons diffuse across
small distances and erase the primordial fluctuations on these scales.
On the other hand, the 21cm photons originate from a resonant transition of
hydrogen atoms and so they trace the gas distribution on all scales. The
gas is expected to be inhomogeneous on all scales down to where pressure
counteracts gravity and smoothes the inhomogeneities through sound
waves. As it turns out, this minimum clumping scale for the gas is many
orders of magnitude smaller than the diffusion scale of the microwave
background photons. Consequently, the 21cm photons can trace the primordial
inhomogeneities with a much finer resolution (i.e. many more independent
pixels) than the microwave background.  Third, the microwave background
fluctuations originated at an early time before the first galaxies formed
in the Universe and so they mainly carry information about the small seed
inhomogeneities that existed in the early Universe. The 21cm fluctuations
probe these inhomogeneities as well as the bubbles of ionized gas around
groups of galaxies (in the form of cavities in the distribution of neutral
hydrogen). Hence the 21cm data could inform us about the initial conditions
of the Universe as well as on the environmental impact of the first
galaxies on their cosmic habitat.

{\it Why is it then that 21cm mapping of the infant Universe was not done
already?}  Detection of the redshifted 21cm signal is
challenging. Low-frequency foreground from radio broadcasting on Earth can
be eliminated by frequency filtering techniques and most importantly by
choosing wisely the observatory site. However, it is impossible to avoid
the fact that relativistic electrons within our Milky-Way galaxy produce
synchrotron radio emission as they gyrate around the Galactic magnetic
field. This produces a radio foreground that is at least a factor of ten
thousand larger than the expected reionization signal. But not all is
lost. By shifting slightly in observed wavelength one is slicing the
hydrogen distribution at different redshifts and hence one is seeing a
different map of its bubble structure, but the synchrotron foreground
remains nearly the same. Theoretical calculations demonstrate that it is
possible to extract the signal from the epoch of reionization by
subtracting the radio images of the sky at slightly different wavelengths.

\begin{figure} 
\plotone{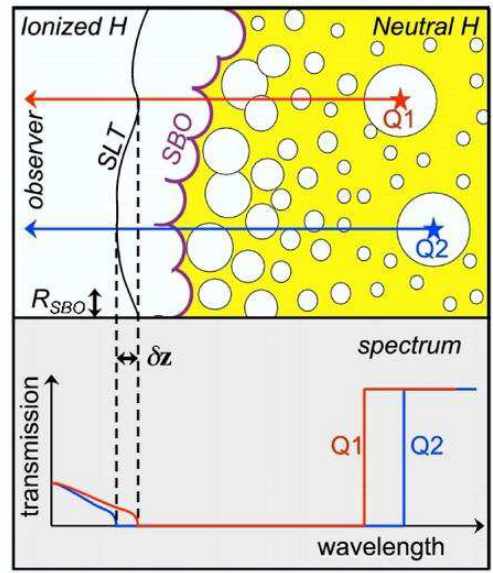} 
\caption{The distances to the observed Surface of Bubble Overlap (SBO)
fluctuate on the sky (from Wyithe \& Loeb 2004). The SBO corresponds to the
first region of diffuse neutral hydrogen {\em observed} along a random
line-of-sight.  It fluctuates across a shell with a minimum width dictated
by the condition that the light crossing time across the characteristic
radius $R_{\rm SBO}$ of ionized bubbles equals the cosmic scatter in their
formation times. After some time delay the ionized cosmic gas becomes
transparent to Ly$\alpha$ photons (which resonate with the transition of
hydrogen from its ground state to the first excited level), resulting in a
second surface, the Surface of Ly$\alpha$ Transmission (SLT).  The upper
panel illustrates how the lines-of-sight towards two quasars (Q1 in red and
Q2 in blue) intersect the SLT with a redshift difference $\delta z$. The
resulting variation in the observed spectrum of the two quasars is shown in
the lower panel.}
\label{ffig1}
\end{figure}

\begin{figure}
\plotone{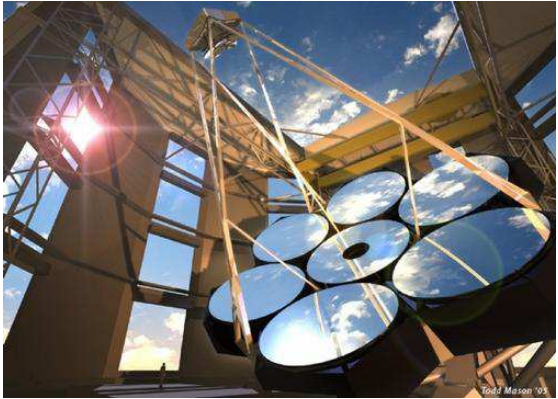}
\caption{Artist conception of the design for one of the future giant
telescopes that could probe the first generation of galaxies from the
ground. The {\it Giant Magellan Telescope} (GMT) will contain seven mirrors
(each 8.4 meter in diameter) and will have the resolving power equivalent
to a 24.5 meter (80 foot) primary mirror. For more details see
http://www.gmto.org/}
\label{gmt}
\end{figure}

In parallel to the search for redshifted 21cm fluctuations, future infrared
telescopes will search directly for the early galaxies that induced some of
these fluctuations.  The next generation of ground-based telescopes will
have a diameter of 24-42 meters (examples include the {\it Giant Magellan
Telescope}, the {\it Thirty Meter Telescope}, and the {\it European
Extremely Large Telescope}). Together with the infrared space telescope
JWST (that will not be affected by the atmospheric opacity and emission),
they will be able to image the first galaxies. Given that these galaxies
also created the ionized bubbles around them, the locations of galaxies
should correlate with bubbles in the neutral hydrogen. Within a decade it
would be possible to explore the environmental influence of individual
galaxies by using both radio and infrared instruments in concert.

\begin{figure}
\centering
\includegraphics[height=3.5cm]{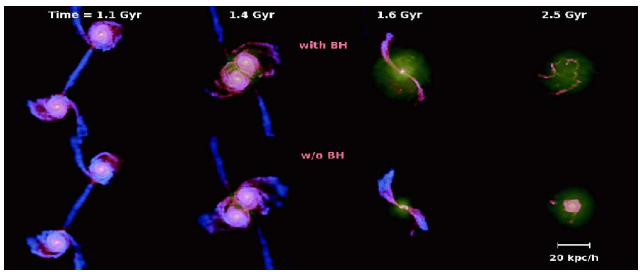}
\caption{Simulation images of a merger of galaxies resulting in quasar
activity that eventually shuts-off the accretion of gas onto the black hole
(from Di Matteo et al. 2005). The upper (lower) panels show a sequence of
snapshots of the gas distribution during a merger with (without) feedback
from a central black hole. The temperature of the gas is color coded.}
\label{merger}
\end{figure}

\begin{figure}
\includegraphics[height=6cm]{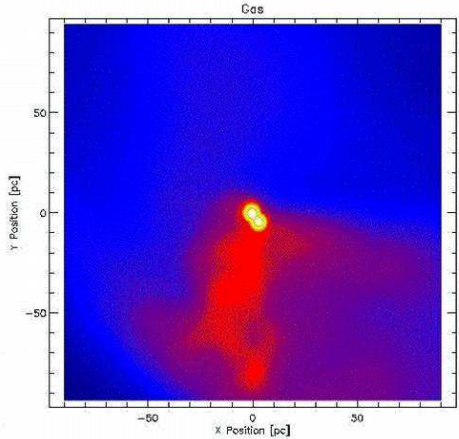}
\caption{Numerical simulation of the collapse of an early dwarf galaxy with
a virial temperature just above the cooling threshold of atomic hydrogen
and no H$_2$ (from Bromm \& Loeb 2003).  The image shows a snapshot of the
gas density distribution 500 million years after the big bang, indicating
the formation of two compact objects near the center of the galaxy with
masses of $2.2\times 10^{6}M_{\odot}$ and $3.1\times 10^{6}M_{\odot}$,
respectively, and radii $<1$ pc. Sub-fragmentation into lower mass clumps
is inhibited because hydrogen atoms cannot cool the gas significantly below
its initial temperature.  These circumstances lead to the formation of
supermassive stars that inevitably collapse and trigger the birth of
supermassive black holes.  The box size is 635 light years across.  }
\label{fig6}
\end{figure}

\section{\bf But the Brightest Sources Across the Universe are not Galaxies!}

As already mentioned, imaging the infant Universe would reveal how the
building blocks of present-day galaxies originated. Of particular interest
is the formation history of massive black holes in the centers of these
galaxies.  It has been recently realized that almost every galaxy in the
present-day Universe hosts a massive black hole at its nucleus. In our own
Milky-Way galaxy, stars were found to zoom around the Galactic center with
speeds of up to ten thousand kilometers per second which could only be
induced by a massive object which is as compact as a black hole. The
nuclear black holes in galaxies are believed to be fed with gas in episodic
events of gas accretion which are triggered by mergers of galaxies.  The
energy released by the accreting gas during these episodes could easily
unbind the gas reservoir from the host galaxy and suppress star formation
within it (see Fig. \ref{merger}). If so, nuclear black holes regulate
their own growth by expelling the gas that feeds them, and in so doing they
also shape the stellar content of their host galaxy.  Such feedback has
been invoked to explain the observed correlation between the masses of
black holes and the properties of the galaxy (most importantly, the depth
of its gravitational potential well) in which they are embedded. During
their episodes of growth, the accreting gas shines much brighter than the
entire galaxy surrounding it and appears as a quasar. Quasars are often a
hundred times more luminous than their host galaxy. Deep observations by
the {\it Sloan Digital Sky Survey} have revealed that quasars with black
holes masses of more than a billion suns already exist in the Universe when
it had only 6\% of its present-age.  These masses are comparable to the
most massive black holes found today. {\it How did such massive black holes
come to exist so early?  Why don't we observe black holes with much higher
masses today?} The answers to these questions could be found by imaging the
early Universe.

\begin{figure}
\begin{center}
\includegraphics[height=4cm]{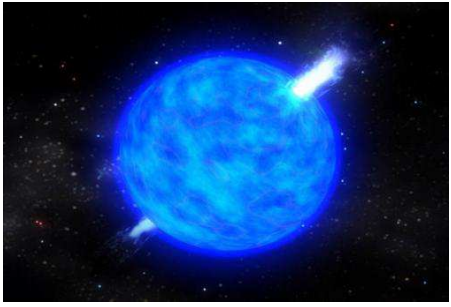}
\end{center}
\caption{Illustration of a long-duration gamma-ray burst in the popular
``collapsar'' model. 
The collapse of the core of a massive star (which lost
its hydrogen envelope) to a black hole generates two opposite jets moving
out at a speed close to the speed of light. The jets drill a hole in the
star and shine brightly towards an observer who happened to be located
within with the collimation cones of the jets. The jets emenating from a
single massive star are so bright that they can be seen across the Universe
out to the epoch when the first stars have formed.  Upcoming observations
by the {\it Swift} satellite will have the sensitivity to reveal whether
the first stars served as progenitors of gamma-ray bursts
(for updates see http://swift.gsfc.nasa.gov/).}
\label{grb}
\end{figure}

Explosions of individual massive stars (known as supernovae) can also
outshine their host galaxies for brief periods of time. The brightest of
these explosions show up as {\it gamma-ray bursts}, namely bright flashes
of high-energy photons followed by afterglows at lower photon
energies. These afterglows can be used to study the first stars, one star
at a time (see Fig. \ref{grb}). Together with quasars they can also be used
as beacons of light that reveal the state of the cosmic gas along the
line-of-sight to them through its imprint of absorption lines on their
spectra. The earliest gamma-ray burst was discovered by the recently
launched {\it Swift} satellite. It originated at the same epoch (redshift
of $z=6.3$) as the earliest quasar (redshift of $z=6.4$), only a billion
years after the big bang. Detection of even earlier gamma-ray bursts will
open a new window into the infant Universe where our origins lie.

\vfil \eject

\bigskip

\noindent
{\underline{\bf Additional Reading}}

\medskip
\noindent
{\bf First Light.} {\it A. Loeb, extensive review (157 pages long) prepared
for the SAAS-Fee winter school, 2006; http://arxiv.org/abs/astro-ph/0603360}

\smallskip
\noindent
{\bf The First Stars.}  {\it V. Bromm and R. Larson, Annual Reviews of
Astronomy and Astrophysics, Vol. 42, pages 79-118, 2004;
http://xxx.lanl.gov/abs/astro-ph/0311019}

\end{document}